# Kinetic Theory of Drag on Objects in Nearly Free Molecular Flow

(28 June 2014)

J.V. Sengers[a,*], Y.-Y. Lin Wang[b], B. Kamgar-Parsi[c], and J.R. Dorfman[a]

[a]Institute for Physical Science and Technology, University of Maryland, College Park, MD 20742-8510, USA

[b]Department of Physics, National Taiwan Normal University, Taipei 116, Taiwan

[c]Mathematics, Computer and Information Research Division, Office of Naval Research, Arlington, VA 22203-1995, USA

ABSTRACT

Using an analogy between the density expansion of the transport coefficients of moderately dense gases and the inverse-Knudsen-number expansion of the drag on objects in nearly free molecular flows, we formulate the collision integrals that determine the first correction term to the free-molecular drag limit. We then show how the procedure can be applied to calculate the drag coefficients of an oriented disc and a sphere as a function of the speed ratio.

* Tel.: +1 301 405 4805; fax: +1 301 314 9404.

*E-mail address*: sengers@umd.edu

## 1 Introduction

The drag force on a solid object moving in a rarefied gas has been and remains a subject of great technological interest [1-3]. An important parameter in the theory of rarefied gas flows is the Knudsen number $Kn$, which is the ratio of the mean free path $l$ of the molecules and a length $R$ that characterizes the size of the object. The limit $Kn \to \infty$ corresponds to the free-molecular flow regime in which the drag is solely determined by the number of individual gas molecules striking the object and the collision dynamics. The limit $Kn \to 0$ corresponds to the continuum regime in which one needs to solve the full nonlinear Boltzmann equation subject to the appropriate boundary conditions. A proper theory for the drag force at arbitrary Knudsen numbers is an interesting and important challenge [3-9]. In the absence of reliable theoretical predictions, one often resorts to empirical correlations [10].

Here we consider the drag coefficient of objects in nearly free molecular flow, where the Knudsen number is large but not infinite. As pointed out by Dorfman *et al.* [11-13], there is a close similarity between the density expansion of transport coefficients of moderately dense gases and an expansion of the drag coefficient of objects around the free-molecular-flow limit in terms of the inverse Knudsen number $Kn^{-1}$. For instance, the viscosity $\mu$ of a moderately dense gas has an expansion in terms of the density $n$ of the form [14]

$$\mu = \mu_0 + \mu_1 n + \mu_2' n^2 \log n + \mu_2'' n^2 + \ldots \tag{1.1}$$

In this expansion the viscosity $\mu_0$ in the dilute-gas limit is determined by uncorrelated binary collisions between the gas molecules, the coefficient $\mu_1$ by correlated collision sequences among three molecules, and $\mu_2'$ by correlated collision sequences among four molecules. The drag coefficient $C_D$ of an object is defined as

$$C_D = \frac{F}{U_K}, \tag{1.2}$$

where $F$ is the magnitude of the force exerted on the object and $U_K$ the incident kinetic energy. In the nearly free-molecular flow regime this drag coefficient has an expansion of the form

$$C_D = C_0 + C_1 Kn^{-1} + C_2' Kn^{-2} \log Kn^{-1} + C_2'' Kn^{-2} + \ldots \tag{1.3}$$

The expressions for the coefficients in the expansion of the drag coefficient can be obtained by applying a Knudsen-number iteration to the solution of the Boltzmann equation in the presence of the object [12, 13]. One then finds that the expressions for the coefficients in (1.3) are related to the same dynamical events that determine the coefficients in (1.1), but with one of the molecules replaced with the foreign object. The expansion (1.3) is valid when the mean free path of the molecules is large compared to the size of the object in all three dimensions. When the mean free path is large compared to the size of the object in two dimensions, but not in the third dimension, for example, in the case of the drag on a cylinder or on a strip, a logarithmic dependence on the inverse Knudsen number already appears in the first correction to the free-molecular-flow limit, so that the expansion for the drag coefficient becomes [13, 15-17]

$$C_D = C_0 + C_1'' Kn^{-1} \log Kn^{-1} + C_2'' Kn^{-1} + ..., \tag{1.4}$$

again in analogy to the logarithmic density expansion of the transport coefficients of a two-dimensional gas [18, 19]. In addition it should be noted that the coefficient $C_1''$ of the term linear in $Kn^{-1}$ in (1.4) depends logarithmically on the flow velocity, so that expansion (1.4) is only valid for finite values of the flow velocity [13]. There exists a similar analogy between the density expansion of the transport properties of moderately dense gases and the density expansion of the friction coefficient of a Brownian particle [20].

The present paper will only deal with the drag on objects whose size is small in all directions compared to the mean free path, so that the first correction to the free-molecular drag force is linear in $Kn^{-1}$ in accordance with (1.3). The specific purpose of the present paper is to formulate the collision integrals that determine the amplitude $C_1$ of the first inverse-Knudsen-number correction to the free-molecular flow drag and then show how they can be evaluated to determine the magnitude of this correction for a disc oriented perpendicular to the flow and for a sphere as representative examples. In principle, our method of calculating the drag coefficient from collision integrals can be applied to objects of any shape and for any interactions of the gas molecules with the solid surface of the object. In practice we shall introduce a number of simplifying approximations:

• The molecules that strike the object do not stick to it, but are re-emitted after a time short compared to the mean free time of the molecules.

• The molecules are re-emitted diffusively with a temperature $T$ corresponding to the temperature of the object, which is assumed to be the same as the temperature of the molecules in the gas stream.

• The molecules in the gas stream have a mass $m$ and interact with short-ranged repulsive forces of range $\sigma$.

• The solid object is convex (non-concave), so that a molecule emitted from the surface of the object cannot strike the object unless it first collides with another molecule.

The drag force on the object not only depends on the Knudsen number, but also on the Mach number $M$, defined as the ratio of the magnitude of the flow velocity $V$ and the sound velocity. Instead of the Mach number, we shall use the speed ratio, which is the ratio of $V$ and the thermal molecular velocity:

$$S = V(m/2k_B T)^{1/2}, \tag{1.5}$$

where $k_B$ is Boltzmann's constant. The speed ratio is directly proportional to the Mach number as $S = M(\gamma/2)^{1/2}$, where $\gamma$ is the ratio of the isobaric and isochoric heat capacities [1].

As in the case of the collision integrals appearing in the density expansion of the transport coefficients of moderately dense gases [21, 22], we find it convenient to represent the molecular collisions and the interaction of the molecules with the solid object by binary collision operators to be defined in Sect. 2. Using then an expansion in terms of these binary-collision operators, we formulate in Sect. 3 the specific collision sequences that contribute to the first correction of the drag force beyond the free-molecular flow limit. The explicit expressions for the relevant collision integrals are presented in Sect. 4. As representative examples, we evaluate these collision integrals for the drag coefficient of a disc in Sect. 5 and for a sphere in Sect. 6. A brief summary of our results is presented in Sect. 7.

## 2 Binary Collision Operators

When the density of the gas is sufficiently small so that the mean free path is large compared to the size of the molecules, we may neglect the probability that two molecular collisions occur simultaneously and the dynamical processes reduce to successive collisions among molecules and the object. We assume that a gas molecule after striking the object with an incoming velocity $\mathbf{v}$, will be re-emitted from the surface with a velocity $\mathbf{v}'$ with a probability density $\eta(\mathbf{v}'|\mathbf{v})$. We assume that all molecules impinging on the object will be re-emitted again within a time interval small compared to the mean free time, so that the transition probability is normalized as $\int d\mathbf{v}' \, \eta(\mathbf{v}'|\mathbf{v}) = 1$. To account for the interaction of a molecule $i$ with the surface of the object, it is convenient to introduce an operator $T(i,X)$ defined as [13]

$$T(i,X) = T^{\mathrm{i}}(i,X) + T^{\mathrm{n}}(i,X) \tag{2.1}$$

with

$$T^{\mathrm{i}}(i,X) = \int d\mathbf{v}'_i \int_{\mathbf{v}_i \cdot \hat{\mathbf{n}} \leq 0} d\mathbf{A} \, \eta(\mathbf{v}'_i|\mathbf{v}_i) \left| \mathbf{v}_i \cdot \hat{\mathbf{n}} \right| \delta^3(\mathbf{r} - \mathbf{R}) \mathcal{R}_R \tag{2.1a}$$

$$T^{\mathrm{n}}(i,X) = - \int_{\mathbf{v}_i \cdot \hat{\mathbf{n}} \leq 0} d\mathbf{A} \, \left| \mathbf{v}_i \cdot \hat{\mathbf{n}} \right| \delta^3(\mathbf{r} - \mathbf{R}) \,. \tag{2.1b}$$

Here it is assumed that the molecule with incoming velocity strikes the surface of the object at position $\mathbf{R}$ on the surface inside the two-dimensional surface element $d\mathbf{A}$. The symbol $\delta^3(\mathbf{r} - \mathbf{R})$ represents a three-dimensional delta function. The operator $\mathcal{R}_R$ in (2.1a) is an operator that transforms the velocity $\mathbf{v}_i$ of the molecule $i$ before colliding with the object into the velocity $\mathbf{v}'_i$ after leaving the object. The vector $\hat{\mathbf{n}}$ is the unit vector in the direction of the outward normal and the condition $\left| \mathbf{v}_i \cdot \hat{\mathbf{n}} \right| \leq 0$ indicates that the molecule strikes the surface from the outside. Just as in the density dependence of the transport properties of gases, one must consider collision sequences with both interacting and non-interacting collisions [22]. The meaning of the operators $T^{\mathrm{i}}(i,X)$ and $T^{\mathrm{n}}(i,X)$ is illustrated in Fig. 1. In this figure the circle represents the solid object and the lines represent the trajectory of a molecule. The operators $T^{\mathrm{i}}(i,X)$ and $T^{\mathrm{n}}(i,X)$ are only different from zero when the molecule impinges on the object. The operator $T^{\mathrm{i}}(i,X)$ transforms the incident velocity $\mathbf{v}_i = \mathbf{v}$ of the molecule into the velocity $\mathbf{v}'_i = \mathbf{v}'$ of the molecule after having been re-emitted from the object with the appropriate probability distribution for $\mathbf{v}'$; hence, it represents the actual path of a molecule in the presence of the object. We refer to this process as an "interacting" collision between the molecule and the object as shown in Fig. 1a. The operator $T^n(i, X)$ requires also that the molecule strikes the object, but it does not change the velocity $\mathbf{v}_i = \mathbf{v}$: hence, it shows the path of the molecule in the absence of the object. We refer to this case as a "non-interacting" collision in which the velocity of the molecule remains equal to the velocity prior to the collision, as shown in Fig. 1b.

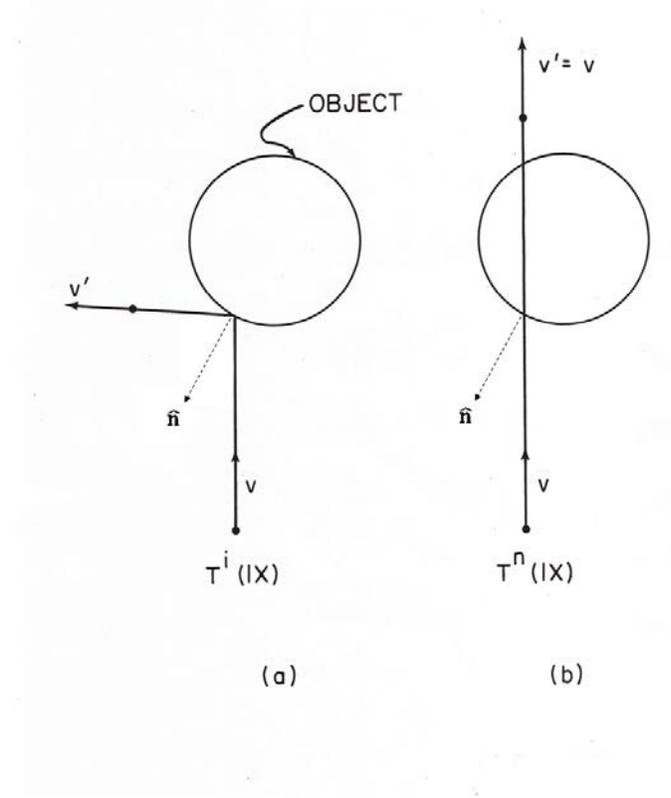

**Fig. 1** Schematic representation of the trajectory of a molecule in the presence of an object.

(a): An "interacting " collision with the object. (b): A "non-interacting" collision with the object.

The operator $T^i(i, X)$ is a generalization of the hard-sphere binary collision operators $T(ij)$ introduced by Ernst *et al.* [23]. They are defined as

$$T(ij) = T^i(ij) + T^n(ij), \tag{2.2}$$

where $T^i(ij)$ corresponds to an interacting collision between two hard sphere molecules $i$ and $j$:

$$T^i(ij) = \sigma^2 \int_{\mathbf{v}_{ij} \cdot \hat{\boldsymbol{\sigma}}_{ij} \leq 0} d\hat{\boldsymbol{\sigma}} \left|\mathbf{v}_{ij} \cdot \hat{\boldsymbol{\sigma}}_{ij}\right| \delta^3(\mathbf{r}_{ij} - \boldsymbol{\sigma}_{ij}) \mathcal{R}_{\sigma_{ij}}, \tag{2.2a}$$

and where $T^n(ij)$ corresponds to a non-interacting collision between two hard sphere molecules:

$$T^n(ij) = -\sigma^2 \int_{\mathbf{v}_{ij}\cdot\hat{\boldsymbol{\sigma}}_{ij}\leq 0} d\hat{\boldsymbol{\sigma}} \left|\mathbf{v}_{ij}\cdot\hat{\boldsymbol{\sigma}}_{ij}\right| \delta^3(\mathbf{r}_{ij} - \boldsymbol{\sigma}_{ij}). \tag{2.2b}$$

Here $\mathbf{v}_{ij} = \mathbf{v}_i - \mathbf{v}_j$, $\mathbf{r}_{ij} = \mathbf{r}_i - \mathbf{r}_j$ are the relative velocity and position of the two molecules, $\sigma$ the diameter of the molecules, and $\boldsymbol{\sigma}_{ij}$ is the collision vector from the center of molecule $j$ to the center of molecule $i$ at time of contact with $\hat{\boldsymbol{\sigma}}_{ij}$ being the corresponding unit vector. The operator $\mathcal{R}_{\sigma_{ij}}$ in (2.2a) transforms the velocities $\mathbf{v}_i, \mathbf{v}_j$ prior to the collision into the velocities $\mathbf{v}'_i, \mathbf{v}'_j$ after the collision:

$$\mathbf{v}'_i = \mathcal{R}_{\sigma_{ij}} \mathbf{v}_i = \mathbf{v}_i - \left(\mathbf{v}_{ij}\cdot\hat{\boldsymbol{\sigma}}_{ij}\right)\hat{\boldsymbol{\sigma}}_{ij}, \qquad \mathbf{v}'_j = \mathcal{R}_{\sigma_{ij}} \mathbf{v}_j = \mathbf{v}_j + \left(\mathbf{v}_{ij}\cdot\hat{\boldsymbol{\sigma}}_{ij}\right)\hat{\boldsymbol{\sigma}}_{ij}. \tag{2.3}$$

The properties of these binary-collision operators have been discussed in detail elsewhere [24, 25]. The explicit expressions (2.2) for the binary collision operators refer to hard sphere molecules. However, the crucial assumption is that for the mean free paths considered we can neglect simultaneous multiple molecular collisions. Hence, we can use this description for a gas of molecules with any finite-ranged repulsive interaction, provided that the hard-sphere cross section in (2.2) is replaced with the actual cross section of the molecules under consideration. Also we do not need to distinguish between $T(ij)$ and $\overline{T}(ij)$, as in the theory of transport properties of moderately dense gases [21, 24], since we assume that the range $\sigma$ is small compared to the mean free path of the molecules in the gas. To account for the time difference between successive collisions, we also introduce a free streaming operator $S^0(\tau)$ such that

$$S^0_\tau \mathbf{r}_1 = \mathbf{r}_1 + \mathbf{v}_1 \tau, \quad S^0_\tau \mathbf{r}_2 = \mathbf{r}_2 + \mathbf{v}_2 \tau, \quad S^0_\tau \mathbf{v}_1 = \mathbf{v}_1, \quad S^0_\tau \mathbf{v}_2 = \mathbf{v}_2. \tag{2.4}$$

## 3  Collision Sequences

We consider the drag force $\mathbf{F}$ on an object in a gas with a molecular distribution of the form

$$f(\mathbf{v}_i; \mathbf{V}) = \left(\frac{m}{2\pi k_B T}\right)^{3/2} \exp\left\{-\left[\frac{m(\mathbf{v}_i - \mathbf{V})^2}{2k_B T}\right]\right\}, \tag{3.1}$$

where the distribution function is expressed in the rest frame of the object with $\mathbf{V}$ being the average velocity of the gas stream relative to the object. As mentioned in the introduction, we assume that the temperature of the solid is the same as that of the gas. For large values of the Knudsen number we write this drag force as

$$\mathbf{F} = \mathbf{F}_0 + \mathbf{F}_1 + \ldots, \tag{3.2}$$

where $\mathbf{F}_0$ is the drag force in the free-molecular flow limit, and where $\mathbf{F}_1$ is the first correction which turns out to be proportional to the inverse Knudsen number. Higher-order terms in the expansion (3.2) for the drag force will not be considered any further in this paper.

A systematic procedure for determining the terms in the expansion (3.2) for the drag force has been presented in an earlier publication [13] and has been further elucidated in a technical report [26]. One obtains

$$\mathbf{F}_0 = -mn \int d\mathbf{r}_1 \int d\mathbf{v}_1 \, f(\mathbf{v}_1; \mathbf{V}) T(1X) \mathbf{v}_1, \tag{3.3}$$

$$\mathbf{F}_1 = -mn^2 \int d\mathbf{r}_1 \int d\mathbf{r}_2 \int d\mathbf{v}_1 \int d\mathbf{v}_2 \, f(\mathbf{v}_1; \mathbf{V}) f(\mathbf{v}_2; \mathbf{V}) \mathbf{B}. \tag{3.4}$$

The drag force $\mathbf{F}_0$ in the free-molecular-flow limit is caused by individual molecules 1 striking the object. The drag force $\mathbf{F}_1$ is caused by interactions of the object with pairs of molecules 1 and 2. One can account for these dynamical events by expanding $\mathbf{B}$ in the integrand of (3.4) in terms of the binary-collision operators defined in the preceding section:

$$\mathbf{B} = \mathbf{B}_3 + \mathbf{B}_4 + \ldots \tag{3.5}$$

with

$$\mathbf{B}_3 = T(1X) S^0 * T(12) S^0 \left[ T(1X) + T(2X) \right] (\mathbf{v}_1 + \mathbf{v}_2), \tag{3.6}$$

$$\mathbf{B}_4 = T(1X) S^0 * T(2X) S^0 * T(12) S^0 \left[ T(1X) + T(2X) \right] (\mathbf{v}_1 + \mathbf{v}_2). \tag{3.7}$$

In these expressions we have introduced as a short-hand notation the convolution product

$$g * h = \lim_{t \to \infty} \int_0^t d\tau \, g(\tau) h(t - \tau) = \lim_{t \to \infty} \int_0^t d\tau' \, g(t - \tau') h(\tau'). \tag{3.8}$$

Taking the limit $t \to \infty$ means that we do not consider any transient effects, but evaluate the drag force in the stationary state. The term $\mathbf{B}_3$ accounts for three successive dynamical events involving the object and two molecules. The term $\mathbf{B}_4$ accounts for four successive dynamical events involving the object and two molecules. The binary-collision expansion associated with the interaction of two molecules and a non-concave object terminates after $\mathbf{B}_4$.

To elucidate the nature of these dynamical events, we use (2.1) and (2.2) to write $\mathbf{B}_3$ more explicitly as

$$\begin{aligned}
\mathbf{B}_3 = &\, T^i(1X)S^0 * T^i(12)S^0 T(1X)\mathbf{v}_1 + & \text{(R1)} \\
&\, T^n(1X)S^0 * T^i(12)S^0 T(1X)\mathbf{v}_1 + & \text{(R2)} \\
&\, T^i(1X)S^0 * T^i(12)S^0 T(2X)\mathbf{v}_2 + & \text{(C1)} \\
&\, T^n(1X)S^0 * T^i(12)S^0 T(2X)\mathbf{v}_2 + & \text{(C2)} \\
&\, T^i(1X)S^0 * T^n(12)S^0 T(2X)\mathbf{v}_2 + & \text{(H1)} \\
&\, T^n(1X)S^0 * T^n(12)S^0 T(2X)\mathbf{v}_2 & \text{(H2)}
\end{aligned} \qquad (3.9)$$

According to (3.9), $\mathbf{B}_3$ can be decomposed into a sum of six terms corresponding to six different types of sequences of three successive collisions among two molecules and the object. In analogy to a previous analysis of successive collisions in the kinetic theory of transport properties [18], we refer to these sequences as R1 and R2 (recollisons), C1 and C2 (cyclic collisions), and H1 and H2 (hypothetical collisions). These collision sequences are represented by the six diagrams in Fig. 2. In these diagrams the lines with labels 1 and 2 indicate trajectories of molecules 1 and 2. The vertical line represents the surface of the object. The molecules traverse their trajectories in the direction indicated by arrows. An interacting collision between the two molecules or between a molecule and the object causes a change in the direction of the trajectories. In a non-interacting collision the molecules continue to proceed in the direction of their prior trajectory; in the diagram we have added a shaded region where we want to indicate the occurrence of a non-interacting collision.

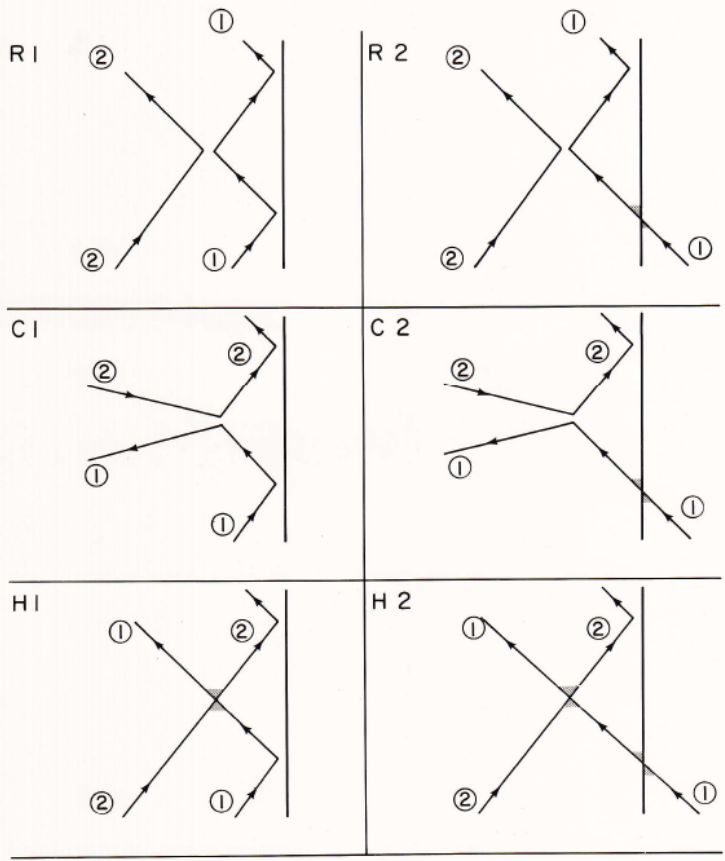

**Fig. 2** Sequences of successive collisions among two molecules and the object

Molecule 1 initially strikes the object and is either reflected by the object (R1, C1, H1) or passes through the object (R2, C2, H2). It then collides with molecule 2 such that molecule 1 either collides with the object again (R1, R2), or causes 2 to collide with the object (C1, C2), or it prevents molecule 2 from colliding with the object (H1, H2). Many of these terms (R$_2$, C$_2$, H$_2$) describe processes that no longer take place, but that are included in the expression (3.6) for $\mathbf{F}_0$, as if they had occurred.

The term $\mathbf{B}_4$, given by (3.8), can similarly be decomposed into a sum of terms corresponding to eight different types of four successive collisions among two molecules and the object. A complete classification of these events can be found elsewhere [16, 26]. However, in practice the contributions of these higher-order dynamical events are very small. Hence, we use the approximation $\mathbf{B} = \mathbf{B}_3$ in (3.5) and do not present in this paper a detailed analysis of the higher-order events that contribute to $\mathbf{B}_4$.

## 4 Collision Integrals for the Drag Force in Nearly Free Molecular Flow

The drag force in the free-molecular-flow regime is determined by (3.3). Using the explicit expression (2.1) for $T(1X)$ we obtain

$$\mathbf{F}_0 = -mn \int d\mathbf{v}_1 \, f(\mathbf{v}_1; \mathbf{V}) \int_{\mathbf{v}_1 \cdot \hat{\mathbf{n}} \leq 0} d\mathbf{A}_1 \, |\mathbf{v}_1 \cdot \hat{\mathbf{n}}_1| \int d\mathbf{v}_1' \, \eta(\mathbf{v}_1' | \mathbf{v}_1)(\mathbf{v}_1' - \mathbf{v}_1). \tag{4.1}$$

The force $\mathbf{F}_1$ to be added in nearly free-molecular flow, given by (3.4), can be decomposed into

$$\mathbf{F}_1 = \mathbf{F}_{R1} + \mathbf{F}_{R2} + \mathbf{F}_{C1} + \mathbf{F}_{C2} + \mathbf{F}_{H1} + \mathbf{F}_{H2}, \tag{4.2}$$

where each term is uniquely related to the corresponding term in (3.9). To specify the integrals determining these contributions, we denote the initial velocities of the molecules by $\mathbf{v}_i$, the velocities after the first collision in the sequence by $\mathbf{v}_i'$, the velocities after the second collision by $\mathbf{v}_i''$, and the velocities after the third collision by $\mathbf{v}_i'''$, as indicated in Fig. 3 for the recollision sequences, in Fig. 4 for the cyclic collision sequences, and in Fig. 5 for the hypothetical collision sequences. In these figures, the circle represents the object, while $\mathbf{R}_1$ and $\mathbf{R}_2$ are the position vectors specifying where a first and a second collision of a molecule with the object occurs..

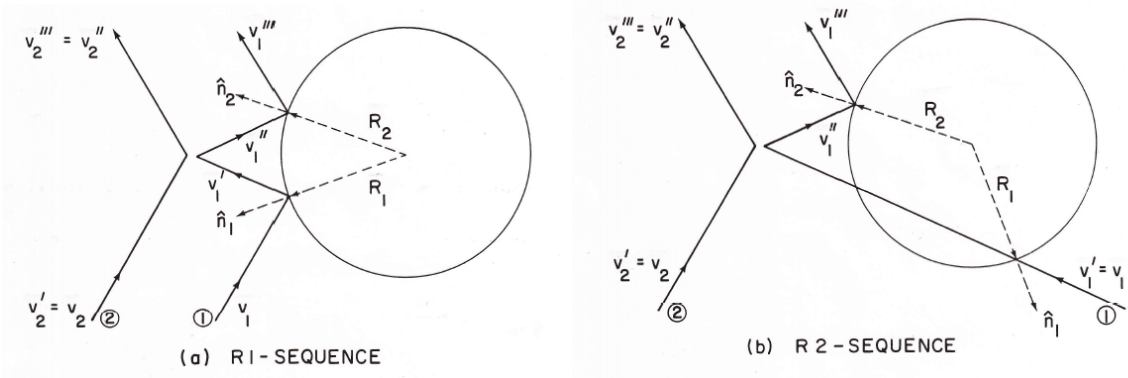

**Fig. 3** Geometric representation of recollision sequences

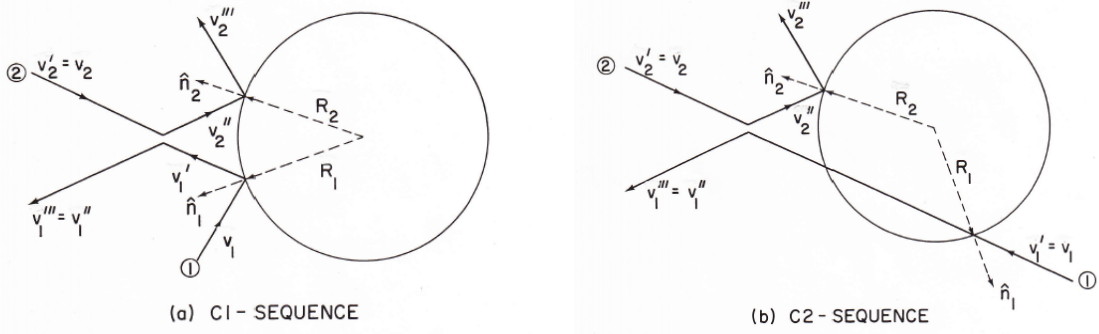

**Fig. 4** Geometric representation of cyclic collision sequences

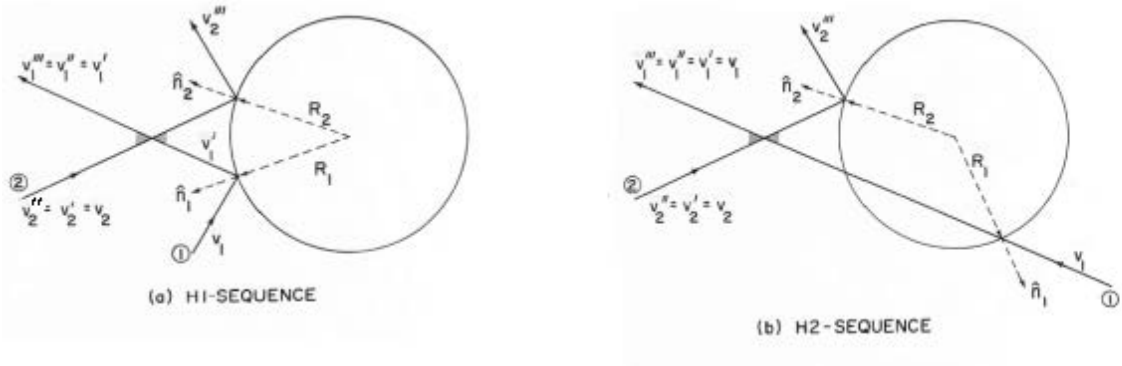

**Fig. 5** Geometric representation of hypothetical collision sequences

Substituting $\mathbf{B} = \mathbf{B}_3$ into (3.4) and using the explicit expressions for the collision operator and for the streaming operator presented in Sect. 3, we obtain

$$\mathbf{F}_{R1} = -mn^2\sigma^2 \iint d\mathbf{v}_1 d\mathbf{v}_2 f(\mathbf{v}_1;\mathbf{V}) f(\mathbf{v}_2;\mathbf{V}) \int_0^\infty d\tau_1 \int_0^\infty d\tau_2$$

$$\int_{\mathbf{v}_1\cdot\hat{\mathbf{n}}\leq 0} d\mathbf{A}_1 \left|\mathbf{v}_1\cdot\hat{\mathbf{n}}_1\right| \int d\mathbf{v}_1' \eta(\mathbf{v}_1'|\mathbf{v}_1) \int_{\mathbf{v}_{12}'\cdot\hat{\sigma}_{12}\leq 0} d\hat{\sigma}_{12} \left|\mathbf{v}_{12}'\cdot\hat{\sigma}_{12}\right| \int_{\mathbf{v}_1''\cdot\hat{\mathbf{n}}_2\leq 0} d\mathbf{A}_2 \left|\mathbf{v}_1''\cdot\hat{\mathbf{n}}_2\right| \qquad (4.3)$$

$$\delta^3\left(\mathbf{R}_1 - \mathbf{R}_2 + \mathbf{v}_1'\tau_1 + \mathbf{v}_1''\tau_2\right) \int d\mathbf{v}_1''' \eta(\mathbf{v}_1'''|\mathbf{v}_1'')(\mathbf{v}_1''' - \mathbf{v}_1'')$$

$$\mathbf{F}_{R2} = +mn^2\sigma^2 \iint d\mathbf{v}_1 d\mathbf{v}_2 f(\mathbf{v}_1;\mathbf{V}) f(\mathbf{v}_2;\mathbf{V}) \int_0^\infty d\tau_1 \int_0^\infty d\tau_2$$

$$\int_{\mathbf{v}_1 \cdot \hat{\mathbf{n}}_1 \leq 0} d\mathbf{A}_1 |\mathbf{v}_1 \cdot \hat{\mathbf{n}}_1| \int_{\mathbf{v}_{12} \cdot \hat{\boldsymbol{\sigma}}_{12} \leq 0} d\hat{\sigma}_{12} |\mathbf{v}_{12} \cdot \hat{\boldsymbol{\sigma}}_{12}| \int_{\mathbf{v}_1'' \cdot \hat{\mathbf{n}}_2 \leq 0} d\mathbf{A}_2 |\mathbf{v}_1'' \cdot \hat{\mathbf{n}}_2| \quad (4.4)$$

$$\delta^3(\mathbf{R}_1 - \mathbf{R}_2 + \mathbf{v}_1 \tau_1 + \mathbf{v}_1'' \tau_2) \int d\mathbf{v}_1''' \eta(\mathbf{v}_1'''|\mathbf{v}_1'')(\mathbf{v}_1''' - \mathbf{v}_1'')$$

$$\mathbf{F}_{C1} = -mn^2\sigma^2 \iint d\mathbf{v}_1 d\mathbf{v}_2 f(\mathbf{v}_1;\mathbf{V}) f(\mathbf{v}_2;\mathbf{V}) \int_0^\infty d\tau_1 \int_0^\infty d\tau_2$$

$$\int_{\mathbf{v}_1 \cdot \hat{\mathbf{n}}_1 \leq 0} d\mathbf{A}_1 |\mathbf{v}_1 \cdot \hat{\mathbf{n}}_1| \int d\mathbf{v}_1' \eta(\mathbf{v}_1'|\mathbf{v}_1) \int_{\mathbf{v}_{12}' \cdot \hat{\boldsymbol{\sigma}}_{12} \leq 0} d\hat{\sigma}_{12} |\mathbf{v}_{12}' \cdot \hat{\boldsymbol{\sigma}}_{12}| \int_{\mathbf{v}_2'' \cdot \hat{\mathbf{n}}_2 \leq 0} d\mathbf{A}_2 |\mathbf{v}_2'' \cdot \hat{\mathbf{n}}_2| \quad (4.5)$$

$$\delta^3(\mathbf{R}_1 - \mathbf{R}_2 + \mathbf{v}_1' \tau_1 + \mathbf{v}_2'' \tau_2) \int d\mathbf{v}_2''' \eta(\mathbf{v}_2'''|\mathbf{v}_2'')(\mathbf{v}_2''' - \mathbf{v}_2'')$$

$$\mathbf{F}_{C2} = +mn^2\sigma^2 \iint d\mathbf{v}_1 d\mathbf{v}_2 f(\mathbf{v}_1;\mathbf{V}) f(\mathbf{v}_2;\mathbf{V}) \int_0^\infty d\tau_1 \int_0^\infty d\tau_2$$

$$\int_{\mathbf{v}_1 \cdot \hat{\mathbf{n}}_1 \leq 0} d\mathbf{A}_1 |\mathbf{v}_1 \cdot \hat{\mathbf{n}}_1| \int_{\mathbf{v}_{12} \cdot \hat{\boldsymbol{\sigma}}_{12} \leq 0} d\hat{\sigma}_{12} |\mathbf{v}_{12} \cdot \hat{\boldsymbol{\sigma}}_{12}| \int_{\mathbf{v}_2'' \cdot \hat{\mathbf{n}}_2 \leq 0} d\mathbf{A}_2 |\mathbf{v}_2'' \cdot \hat{\mathbf{n}}_2| \quad (4.6)$$

$$\delta^3(\mathbf{R}_1 - \mathbf{R}_2 + \mathbf{v}_1 \tau_1 + \mathbf{v}_2'' \tau_2) \int d\mathbf{v}_2''' \eta(\mathbf{v}_2'''|\mathbf{v}_2'')(\mathbf{v}_2''' - \mathbf{v}_2'')$$

$$\mathbf{F}_{H1} = +m^2n\sigma^2 \iint d\mathbf{v}_1 d\mathbf{v}_2 f(\mathbf{v}_1;\mathbf{V}) f(\mathbf{v}_2;\mathbf{V}) \int_0^\infty d\tau_1 \int_0^\infty d\tau_2$$

$$\int_{\mathbf{v}_1 \cdot \hat{\mathbf{n}}_1 \leq 0} d\mathbf{A}_1 |\mathbf{v}_1 \cdot \hat{\mathbf{n}}_1| \int d\mathbf{v}_1' \eta(\mathbf{v}_1'|\mathbf{v}_1) \int_{\mathbf{v}_{12}' \cdot \hat{\boldsymbol{\sigma}}_{12} \leq 0} d\hat{\sigma}_{12} |\mathbf{v}_{12}' \cdot \hat{\boldsymbol{\sigma}}_{12}| \int_{\mathbf{v}_2 \cdot \hat{\mathbf{n}}_2 \leq 0} d\mathbf{A}_2 |\mathbf{v}_2 \cdot \hat{\mathbf{n}}_2| \quad (4.7)$$

$$\delta^3(\mathbf{R}_1 - \mathbf{R}_2 + \mathbf{v}_1' \tau_1 + \mathbf{v}_2 \tau_2) \int d\mathbf{v}_2''' \eta(\mathbf{v}_2'''|\mathbf{v}_2'')(\mathbf{v}_2''' - \mathbf{v}_2'')$$

$$\mathbf{F}_{H2} = -mn^2\sigma^2 \iint d\mathbf{v}_1 d\mathbf{v}_2 f(\mathbf{v}_1;\mathbf{V}) f(\mathbf{v}_2;\mathbf{V}) \int_0^\infty d\tau_1 \int_0^\infty d\tau_2$$

$$\int_{\mathbf{v}_1 \cdot \hat{\mathbf{n}}_1 \leq 0} d\mathbf{A}_1 |\mathbf{v}_1 \cdot \hat{\mathbf{n}}_1| \int_{\mathbf{v}_{12} \cdot \hat{\boldsymbol{\sigma}}_{12} \leq 0} d\hat{\sigma}_{12} |\mathbf{v}_{12} \cdot \hat{\boldsymbol{\sigma}}_{12}| \int_{\mathbf{v}_2 \cdot \hat{\mathbf{n}}_2 \leq 0} d\mathbf{A}_2 |\mathbf{v}_2 \cdot \hat{\mathbf{n}}_2| \quad (4.8)$$

$$\delta^3(\mathbf{R}_1 - \mathbf{R}_2 + \mathbf{v}_1 \tau_1 + \mathbf{v}_2 \tau_2) \int d\mathbf{v}_2''' \eta(\mathbf{v}_2'''|\mathbf{v}_2'')(\mathbf{v}_2''' - \mathbf{v}_2'')$$

In these equations $d\mathbf{A}_1$ and $d\mathbf{A}_2$ are the surface elements at the positions $\mathbf{R}_1$ and $\mathbf{R}_2$, respectively. The time $\tau_1$ is the time between the first and the second collision and the time $\tau_2$ is the time between the second and the third collision. In addition we note the following simplifications for the individual collision sequences: $\mathbf{v}_2' = \mathbf{v}_2, \mathbf{v}_2''' = \mathbf{v}_2''$ in R1, $\mathbf{v}_1' = \mathbf{v}_1, \mathbf{v}_2' = \mathbf{v}_2, \mathbf{v}_2''' = \mathbf{v}_2''$ in R2, $\mathbf{v}_2' = \mathbf{v}_2, \mathbf{v}_1''' = \mathbf{v}_1''$ in C1, $\mathbf{v}_1' = \mathbf{v}_1, \mathbf{v}_2' = \mathbf{v}_2, \mathbf{v}_1''' = \mathbf{v}_1''$ in C2, $\mathbf{v}_2'' = \mathbf{v}_2' = \mathbf{v}_2, \mathbf{v}_1''' = \mathbf{v}_1'' = \mathbf{v}_1'$ in H1, $\mathbf{v}_2'' = \mathbf{v}_2' = \mathbf{v}_2, \mathbf{v}_1''' = \mathbf{v}_1'' = \mathbf{v}_1' = \mathbf{v}_1$. If the molecules of the gas are approximated as hard spheres, a major simplification occurs, since one can prove that the recollisons and cyclic collisions then yield identical contributions

$$\mathbf{F}_{C1} = \mathbf{F}_{R1}, \quad \mathbf{F}_{C2} = \mathbf{F}_{R2}. \tag{4.9}$$

A proof of this theorem can be found elsewhere [26].

In conclusion, we may approximate the first inverse-Knudsen-number correction to the drag force by

$$\mathbf{F}_1 = \mathbf{F}_{H1} + \mathbf{F}_{H2} + 2\left(\mathbf{F}_{R1} + \mathbf{F}_{R2}\right) \tag{4.10}$$

where $\mathbf{F}_{R1}$ and $\mathbf{F}_{R2}$ are given by (4.3) and (4.4), and $\mathbf{F}_{H1}$ and $\mathbf{F}_{H2}$ are given by (4.7) and (4.8). The term $\mathbf{F}_{H1}$ represents a "loss" term which accounts for a situation where molecules emitted from the object prevent another molecule from striking the object. The term $2\mathbf{F}_{R1}$ represents a "gain" term, since molecules emitted from the object cause some other molecules to strike the object. The terms $\mathbf{F}_{H2}$ and $2\mathbf{F}_{R2}$ account for a perturbation to the free molecular flow drags, since the object may interrupt the trajectories of molecules, thus preventing some collisions to occur which were included in $\mathbf{F}_0$. The terms $\mathbf{F}_{H2}$ and $2\mathbf{F}_{R2}$ vanish in the infinite-Mach-number limit [16], but need to be included in determining the drag force at low speed ratios.

The expressions (4.3) – (4.9) are valid for velocity distributions of the gas molecules reflected from the object provided that they are re-emitted within a time short compared to the mean free time of the molecules. In applying the theory we shall assume that the molecules are reflected diffusively from the surface with the same temperature $T$ as that of the distribution (3.1) of the incoming molecules:

$$\eta(\mathbf{v}'|\mathbf{v}) = \eta(\mathbf{v}') = \frac{1}{2\pi}\left(\frac{m}{k_B T}\right)^2 \left(\mathbf{v}' \cdot \hat{\mathbf{n}}\right) \exp\left\{-\left[\frac{mv'^2}{2k_B T}\right]\right\} \Theta\left(\mathbf{v}' \cdot \hat{\mathbf{n}}\right), \tag{4.11}$$

where $\Theta(x)$ is a Heaviside function, defined as $\Theta(x) = 1$ for $x \geq 0$ and $\Theta(x) = 0$ for $x \leq 0$.

## 5 Drag Coefficient of a Disc in Nearly Free Molecular Flow

5.1 Collision integrals for drag on a disc

We consider a flat disc with radius $R$ oriented with its face perpendicular to a gas stream with flow velocity $\mathbf{V}$ as schematically shown in Fig. 6. The radius of the disc is small compared to the mean free path. In order to evaluate the collision integrals for the drag coefficient, it is convenient to introduce dimensionless quantities. For this purpose we measure all distances in terms of the disc radius $R$ and all velocities in terms of the thermal velocity $(2k_B T/m)^{1/2}$ of the molecules. We thus define

$$\mathbf{w}_i = \mathbf{v}_i \left( \frac{m}{2k_B T} \right), \qquad \mathbf{S} = \mathbf{V} \left( \frac{m}{2k_B T} \right)^{1/2}, \qquad \tau^* = \frac{\tau}{R} \left( \frac{2k_B T}{m} \right)^{1/2}. \tag{5.1}$$

The surface element of the disc can be written as $d\mathbf{A} = R^2 d\mathbf{r}$, where $\mathbf{r}$ is now a dimensionless two-dimensional vector in the surface plane of the disc ($0 \le r \le 1$). In addition, we introduce a dimensionless distribution function $f^*(\mathbf{w}_i; \mathbf{S})$ and a dimensionless transition probability density $\eta^*(\mathbf{w}'_i)$:

$$f^*(\mathbf{w}_i; \mathbf{S}) = \frac{1}{\pi^{3/2}} \exp\{-(\mathbf{w}_i - \mathbf{S})^2\}, \tag{5.2}$$

$$\eta^*(\mathbf{w}'_i) = \frac{2}{\pi} (\mathbf{w}'_i \cdot \hat{\mathbf{n}}) e^{-w'^2_i} \Theta(\mathbf{w}'_i \cdot \hat{\mathbf{n}}), \tag{5.3}$$

where $\hat{\mathbf{n}}$ is the normal vector at the two surfaces of the disc.

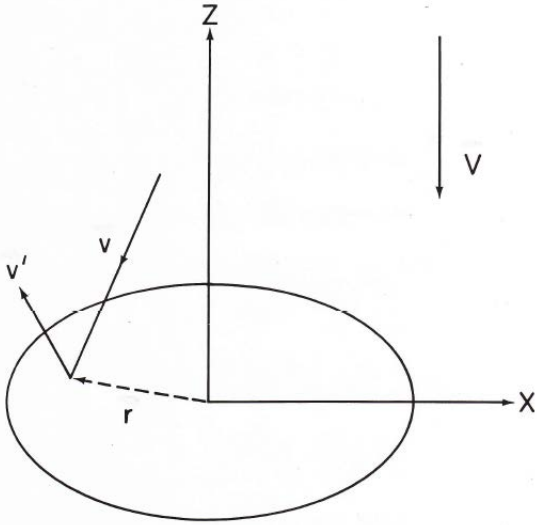

**Fig. 6** Disc in a gas stream with velocity $\mathbf{V}$

The kinetic energy of the gas stream is $U_K = \frac{1}{2} nmV^2 \pi R^2$, so that in accordance with (1.2)

$$C_D = \frac{2F}{nmV^2 \pi R^2}. \tag{5.4}$$

As a result of symmetry, the drag force has only a non-vanishing component in the direction of the flow velocity $\mathbf{V}$, so that $F = \mathbf{F} \cdot \hat{\mathbf{V}}$. For a gas of hard-sphere molecules the Knudsen number may be defined as

$$Kn^{-1} = \sqrt{2}\pi n\sigma^2 R. \tag{5.5}$$

The drag coefficient $C_D$ in the nearly free-molecular-flow regime can be written as

$$C_D = C_0 + C_1 Kn^{-1} + \ldots \tag{5.6}$$

The drag coefficient $C_0$ in the free-molecular-flow limit follows from (4.1)

$$C_0 = -\frac{2}{\pi S^2} \int d\mathbf{w}\, f^*(\mathbf{w};\mathbf{S}) \int_{\mathbf{w}\cdot\hat{\mathbf{n}}\leq 0} d\mathbf{r}\, |\mathbf{w}\cdot\hat{\mathbf{n}}| \int d\mathbf{w}'\, \eta^*(\mathbf{w}')(\mathbf{w}'-\mathbf{w})\cdot\hat{\mathbf{S}}. \tag{5.7}$$

The coefficient $C_1$ of the first inverse-Knudsen-number correction can be decomposed in analogy to (4.10):

$$C_1 = C_{H1} + C_{H2} + 2(C_{R1} + C_{R2}). \tag{5.8}$$

The explicit expressions for the individual terms in (5.8) follow from (4.3), (4.4), (4.7), and (4.8):

$$C_{H1} = +\frac{\sqrt{2}}{\pi^2 S^2} \iint d\mathbf{w}_1 d\mathbf{w}_2\, f^*(\mathbf{w}_1;\mathbf{S}) f^*(\mathbf{w}_2;\mathbf{S}) \int_0^\infty d\tau_1^* \int_0^\infty d\tau_2^*$$
$$\int_{\mathbf{w}_1\cdot\hat{\mathbf{n}}_1\leq 0} d\mathbf{r}_1\, |\mathbf{w}_1\cdot\hat{\mathbf{n}}_1| \int d\mathbf{w}_1^*\, \eta^*(\mathbf{w}_1^*) \int_{\mathbf{w}_{12}'\cdot\hat{\sigma}_{12}\leq 0} d\hat{\sigma}_{12}\, |\mathbf{w}_{12}'\cdot\hat{\sigma}_{12}| \int_{\mathbf{w}_2\cdot\hat{\mathbf{n}}_2\leq 0} d\mathbf{r}_2\, |\mathbf{w}_2\cdot\hat{\mathbf{n}}_2| \tag{5.9}$$
$$\delta^3\left(\mathbf{r}_1 - \mathbf{r}_2 + \mathbf{w}_1'\tau_1^* + \mathbf{w}_2\tau_2^*\right) \int d\mathbf{w}_2'''\, \eta^*(\mathbf{w}_2''')(\mathbf{w}_2'''-\mathbf{w}_2)\cdot\hat{\mathbf{S}}$$

$$C_{H2} = -\frac{\sqrt{2}}{\pi^2 S^2} \iint d\mathbf{w}_1 d\mathbf{w}_2\, f^*(\mathbf{w}_1;\mathbf{S}) f^*(\mathbf{w}_2;\mathbf{S}) \int_0^\infty d\tau_1^* \int_0^\infty d\tau_2^*$$
$$\int_{\mathbf{w}_1\cdot\hat{\mathbf{n}}_1\leq 0} d\mathbf{r}_1\, |\mathbf{w}_1\cdot\hat{\mathbf{n}}_1| \int_{\mathbf{w}_{12}'\cdot\hat{\sigma}_{12}\leq 0} d\hat{\sigma}_{12}\, |\mathbf{w}_{12}\cdot\hat{\sigma}_{12}| \int_{\mathbf{w}_2\cdot\hat{\mathbf{n}}_2\leq 0} d\mathbf{r}_2\, |\mathbf{w}_2\cdot\hat{\mathbf{n}}_2| \tag{5.10}$$
$$\delta^3\left(\mathbf{r}_1 - \mathbf{r}_2 + \mathbf{w}_1\tau_1^* + \mathbf{w}_2\tau_2^*\right) \int d\mathbf{w}_2'''\, \eta^*(\mathbf{w}_2''')(\mathbf{w}_2'''-\mathbf{w}_2)\cdot\hat{\mathbf{S}}$$

$$C_{R1} = -\frac{\sqrt{2}}{\pi^2 S^2} \iint d\mathbf{w}_1 d\mathbf{w}_2\, f^*(\mathbf{w}_1;\mathbf{S}) f^*(\mathbf{w}_2;\mathbf{S}) \int_0^\infty d\tau_1^* \int_0^\infty d\tau_2^*$$
$$\int_{\mathbf{w}_1\cdot\hat{\mathbf{n}}_1\leq 0} d\mathbf{r}_1\, |\mathbf{w}_1\cdot\hat{\mathbf{n}}_1| \int d\mathbf{w}_1'\, \eta^*(\mathbf{w}_1') \int_{\mathbf{w}_{12}'\cdot\hat{\sigma}_{12}\leq 0} d\hat{\sigma}_{12}\, |\mathbf{w}_{12}'\cdot\hat{\sigma}_{12}| \int_{\mathbf{w}_1''\cdot\hat{\mathbf{n}}_2\leq 0} d\mathbf{r}_2\, |\mathbf{w}_1''\cdot\hat{\mathbf{n}}_2| \tag{5.11}$$
$$\delta^3\left(\mathbf{r}_1 - \mathbf{r}_2 + \mathbf{w}_1'\tau_1^* + \mathbf{w}_1''\tau_2^*\right) \int d\mathbf{w}_1'''\, \eta^*(\mathbf{w}_1''')(\mathbf{w}_1'''-\mathbf{w}_1'')\cdot\hat{\mathbf{S}}$$

$$C_{R2} = +\frac{\sqrt{2}}{\pi^2 S^2} \iint d\mathbf{w}_1 d\mathbf{w}_2 f^*(\mathbf{w}_1;\mathbf{S}) f^*(\mathbf{w}_2;\mathbf{S}) \int_0^\infty d\tau_1^* \int_0^\infty d\tau_2^*$$

$$\int_{\mathbf{w}_1\cdot\hat{\mathbf{n}}_1\le 0} d\mathbf{r}_1 |\mathbf{w}_1\cdot\hat{\mathbf{n}}_1| \int_{\mathbf{w}_{12}\cdot\hat{\boldsymbol{\sigma}}_{12}\le 0} d\hat{\boldsymbol{\sigma}}_{12} |\mathbf{w}_{12}\cdot\hat{\boldsymbol{\sigma}}_{12}| \int_{\mathbf{w}_1''\cdot\hat{\mathbf{n}}_2\le 0} d\mathbf{r}_2 |\mathbf{w}_1''\cdot\hat{\mathbf{n}}_2| \qquad(5.12)$$

$$\delta^3\left(\mathbf{r}_1-\mathbf{r}_2+\mathbf{w}_1\tau_1^*+\mathbf{w}_1''\tau_2^*\right) \int d\mathbf{w}_1''' \eta^*(\mathbf{w}_1'')(\mathbf{w}_1'''-\mathbf{w}_1'')\cdot\hat{\mathbf{S}}$$

The drag coefficient $C_D$ is a function of the speed ratio $S$. For a disc $C_D = C_D^+ + C_D^-$, where $C_D^+$ accounts for the force exerted on the front side of the disc and $C_D^-$ accounts for the force at the back side of the disc. The two contributions are interrelated by $C_D^-(+S) = -C_D^+(-S)$.

5.2 Drag coefficient of a disc in the free-molecular-flow limit

The expression (5.7) for $C_0$ can be readily evaluated analytically and one obtains

$$C_0(S) = \frac{2}{S\sqrt{\pi}}\left[e^{-S^2} + \sqrt{\pi}\left(\frac{1}{2S}+S\right)\mathrm{erf}(S) + \frac{\pi}{2}\right] \qquad(5.13)$$

with $\mathrm{erf}(S) \equiv (2/\sqrt{\pi})\int_0^S e^{-t^2} dt$. In the low- and high-speed limits the drag coefficient $C_0$ of a disc reduces to

$$\lim_{S\to 0} C_0(S) = \frac{(4+\pi)}{\sqrt{\pi}S} \square \frac{4.029}{S}, \qquad(5.14)$$

$$\lim_{S\to\infty} C_0(S) = 2. \qquad(5.15)$$

Thus at low speed ratios $C_0(S)$ becomes inversely proportional to the speed ratio $S$, while in the high-speed limit $C_0(S)$ becomes independent of $S$. On comparing with (5.4) we see that at low velocities the drag force $F$ is proportional to the stream velocity as is to be expected, while at high velocities the drag force increases with the square of the stream velocity. The dependence of $C_0(S)$ on the speed ratio $S$ is shown in Fig. 7. Equation (5.13) is in agreement with the result obtained by previous investigators [27].

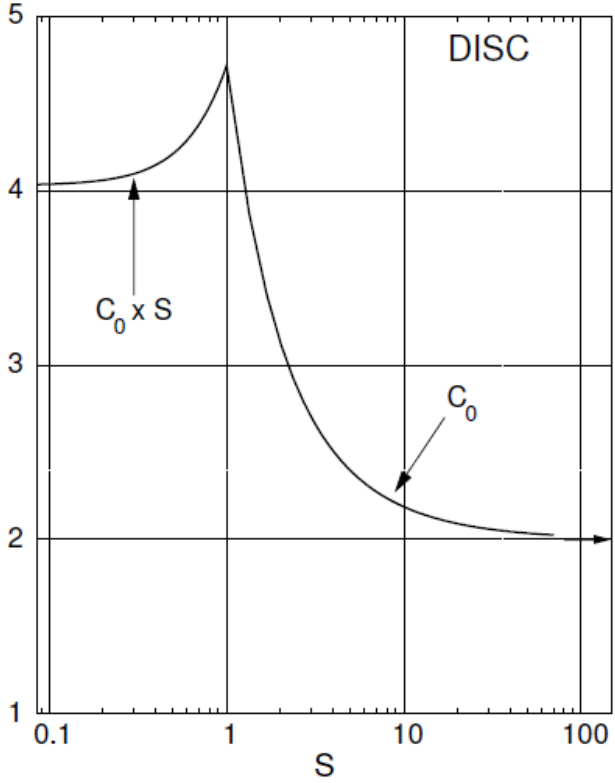

**Fig. 7** Drag coefficient $C_0$ of a disc in free molecular flow as function of the speed ratio $S$

5.3 Drag coefficient of a disc in nearly free-molecular flow

In the nearly free-molecular flow regime the drag coefficient can be represented by

$$C_D = C_0 \left[1 + \frac{C_1}{C_0} Kn^{-1}\right] = C_0 \left[1 + \frac{(C_H + 2C_R)}{C_0} Kn^{-1}\right], \tag{5.16}$$

where $C_H = C_{H1} + C_{H2}$ and $C_R = C_{R1} + C_{R2}$, given by (5.9) - (5.11).

The evaluation of the collision integral becomes simple for $S \gg 1$ to be referred to as the high-speed limit or (cold) wall beam limit. In the beam limit all molecules are moving parallel with a velocity equal to $\mathbf{S}$. Then the distribution functions $f^*(\mathbf{w}_i; \mathbf{S})$ can be approximated by

$$f^*(\mathbf{w}_i; \mathbf{S}) = \delta^3(\mathbf{w}_i - \mathbf{S}). \tag{5.17}$$

Moreover, in the beam limit, the drag force completely originates from collisions at the front surface of the disc. Also in the beam limit two molecules will never collide with each other unless at least one of them interacts with the object, so that the contributions $\lim_{S \to \infty} C_{H2}(S)$ and $\lim_{S \to \infty} C_{R2}(S)$ vanish in this limit. In this limit the collision integrals (5.9) and (5.11) can be readily evaluated analytically and we obtain in the limit of high speed ratios

$$\lim_{S \to \infty} C_{H1}(S) = -\frac{8}{3}\sqrt{\frac{2}{\pi}} S \simeq -2.128 S, \tag{5.18}$$

$$\lim_{S \to \infty} C_{R1}(S) = +\frac{16}{15}\sqrt{\frac{2}{\pi}} S \simeq +0.851 S, \tag{5.19}$$

so that

$$\lim_{S \to \infty} C_1(S) = \lim_{S \to \infty}\{C_{H1}(S) + 2C_{R1}(S)\} = -\frac{8}{15}\sqrt{\frac{2}{\pi}} S \simeq -0.426 S \tag{5.20}$$

The first iterate to the drag coefficient of a disc in a gas of hard spheres in the high-speed limit has also been evaluated by Willis [28]. As quoted by Keel *et al*. [29], he obtained in terms of the variables adopted in this paper $\lim_{S \to \infty} C_1(S) = -0.43 S$ in excellent agreement with the value obtained by us. We note that the coefficient $C_1(S)$ of the first inverse-Knudsen-number correction to the drag coefficient is negative. The physical reason is that the hypothetical collision sequences yield the dominant contribution, as can be seen by comparing (5.18) and (5.19).

Another interesting limit is $S \ll 1$. In the low-speed limit we can expand the Maxwell distribution (5.2) in a Taylor series around $S = 0$

$$f^*(\mathbf{w}_1; \mathbf{S}) f^*(\mathbf{w}_2; \mathbf{S}) \simeq \frac{1}{\pi^3} e^{-(w_1^2 + w_2^2)} [1 - 2S(w_{1z} + w_{2z})], \tag{5.21}$$

where $w_{iz}$ is the component of $\mathbf{w}_i$ in the Z direction indicated in Fig. 6. The zeroth-order term in (5.21) does not contribute to the drag force. In this approximation one can perform the integration over $\mathbf{w}$ in the integrals (5.9) and (5.11) for $C_{H1}$ and $C_{R1}$, resp. After this minor simplification all collision integrals were evaluated numerically by a Monte Carlo method. That is, the integrals were estimated by averaging the integrands over a set of $N$ random points selected in the integration region according to a suitable predetermined probability density function. For this purpose we employed the same method and subroutines that were previously used to calculate the three-particle collision integrals in the density expansion of the transport properties of a gas of hard spheres [22]. The numerical results, together with their estimated standard deviations, obtained by estimating the integrands over 50,000 points, are presented in Table 1.

**Table 1** Coefficient $C_1$ of the first correction to the drag of a disc in the low-speed limit

$$\left[\begin{array}{l} \lim_{S\to 0} C_{H1}(S) = -(3.94\pm 0.13)S^{-1} \\ \lim_{S\to 0} C_{H2}(S) = +(0.23\pm 0.03)S^{-1} \\ \lim_{S\to 0} C_{H}(S) = \lim_{x\to\infty}\{C_{H1}(S)+C_{H2}(S)\} = -(3.71\pm 0.10)S^{-1} \end{array}\right.$$

$$\left[\begin{array}{l} \lim_{S\to 0} C_{R1}(S) = +(1.14\pm 0.07)S^{-1} \\ \lim_{S\to 0} C_{R2}(S) = -(0.05\pm 0.02)S^{-1} \\ \lim_{S\to 0} C_{R}(S) = \lim_{S\to 0}\{C_{R1}(S)+C_{R2}(S)\} = +(1.10\pm 0.08)S^{-1} \end{array}\right.$$

$$\lim_{S\to 0} C_1(S) = \lim_{S\to 0}\{C_H(S)+2C_R(S)\} = -(1.51\pm 0.08)S^{-1}$$

To obtain the drag coefficient at arbitrary values of the speed ratio $S$, we need to retain the full expression (5.2) in the collision integrals (5.9) – (5.12). For various values of the speed ratio $S$, the collision integrals were again evaluated numerically. The results obtained by averaging over 40,000 random points are presented in Table 2.

**Table 2** Drag coefficient of a disc in the nearly free molecular flow regime as a function of the speed ratio $S$.

$$C_D = C_0 \left[ 1 + \frac{C_1}{C_0} Kn^{-1} \right] = C_0 \left[ 1 + \left( \frac{C_H + 2C_R}{C_0} \right) Kn^{-1} \right]$$

| $S$ | $C_0$ | $C_H / C_0$ | $C_R / C_0$ | $C_1 / C_0$ |
|---|---|---|---|---|
| 0    | 4.03 $S^{-1}$ | −0.92 ± 0.02 | +0.27 ± 0.02 | −0.38 ± 0.02 |
| 0.1  | 4.04 $S^{-1}$ | −0.91 ± 0.02 | +0.28 ± 0.02 | −0.36 ± 0.05 |
| 0.5  | 4.21 $S^{-1}$ | −1.14 ± 0.03 | +0.33 ± 0.03 | −0.48 ± 0.06 |
| 1.0  | 4.72 $S^{-1}$ | −1.68 ± 0.03 | +0.45 ± 0.02 | −0.77 ± 0.05 |
| 1.0  | 4.72 | − (1.68 ± 0.03) $S$ | + (0.45 ± 0.02) $S$ | − (0.77 ± 0.05) $S$ |
| 2.0  | 3.14 | − (1.47 ± 0.01) $S$ | + (0.44 ± 0.01) $S$ | − (0.60 ± 0.02) $S$ |
| 3.0  | 2.59 | − (1.29 ± 0.01) $S$ | + (0.43 ± 0.01) $S$ | − (0.43 ± 0.01) $S$ |
| 5.0  | 2.39 | − (1.25 ± 0.01) $S$ | + (0.43 ± 0.01) $S$ | − (0.38 ± 0.01) $S$ |
| 7.5  | 2.25 | − (1.19 ± 0.01) $S$ | + (0.43 ± 0.01) $S$ | − (0.34 ± 0.01) $S$ |
| 10   | 2.19 | − (1.16 ± 0.01) $S$ | + (0.42 ± 0.01) $S$ | − (0.32 ± 0.01) $S$ |
| 15   | 2.12 | − (1.14 ± 0.01) $S$ | + (0.42 ± 0.01) $S$ | − (0.29 ± 0.01) $S$ |
| 20   | 2.09 | − (1.11 ± 0.01) $S$ | + (0.43 ± 0.01) $S$ | − (0.26 ± 0.01) $S$ |
| 30   | 2.06 | − (1.10 ± 0.01) $S$ | + (0.43 ± 0.01) $S$ | − (0.26 ± 0.01) $S$ |
| 40   | 2.04 | − (1.09 ± 0.01) $S$ | + (0.43 ± 0.01) $S$ | − (0.23 ± 0.01) $S$ |
| 50   | 2.04 | − (1.08 ± 0.01) $S$ | + (0.42 ± 0.01) $S$ | − (0.23 ± 0.01) $S$ |
| ∞    | 2.00 | −1.064 $S$ | +0.426 $S$ | −0.213 $S$ |

Note: $Kn^{-1} = \sqrt{2}\pi n \sigma^2 R$

It turns out that the contributions $C_{H2}$ and $C_{R2}$ can be neglected for $S \geq 2$ [26]. Rather than the coefficient $C_1$ itself, we give in Table 2 the relative contributions $C_H / C_0$ and $C_R / C_0$ to $C_1 / C_0$. The numerical results for $S = 0.1$ are in satisfactory agreement with the low-velocity values in Table 1, while at high velocities the results do indeed approach the exact values given by (5.18) and (5.19). The dependence of $C_1 / C_0$ on the speed ratio $S$ is shown in Fig. 8. In nearly free-molecular flow the drag coefficient $C_D$ decreases with decreasing Knudsen number. As noted earlier for the cold beam limit, the sign of the effect is determined by the contribution $C_H \square C_{H1}$, i.e., by the phenomenon that reflected molecules prevent incident molecules from striking the object. The dependence of the drag coefficient on $S$ is quite different whether the speed ratio is smaller or larger than unity. At low velocity the drag force is proportional to the speed ratio $S$. Hence, at small velocities both $C_0$ and $C_1$ become inversely proportional to the speed ratio and the ratio $C_1 / C_0$ becomes independent of $S$. On the other hand, at large velocities the free molecular drag force varies with the square of the stream velocity, while the first inverse Knudsen-number correction to the drag force varies with the third power of the stream velocity; thus the ratio $C_1 / C_0$ becomes proportional to the speed ratio $S$.

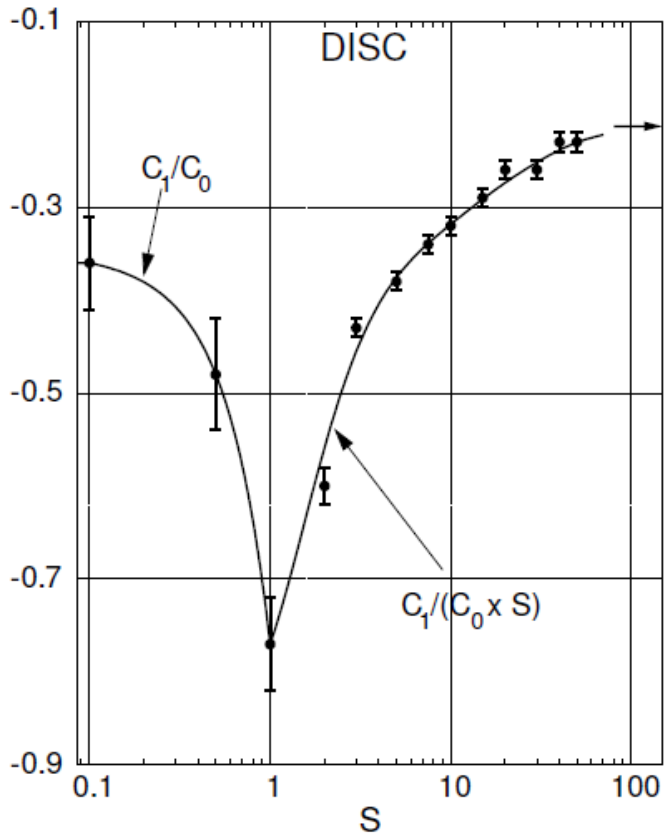

**Fig. 8** Coefficient $C_1 / C_0$ of the first inverse-Knudsen number correction for a disc as a function of the speed ratio $S$. Symbols with error bars represent actual numerical data; curves indicate a smooth interpolation of numerical data.

Experimental data for a disc at normal incidence in a hypersonic nearly free molecular stream of nitrogen molecules have been reported by Keel *et al.* [29]. The experimental results were obtained for reduced wall temperatures $T_w^\star = 0.70$ and $T_w^\star = 0.34$, where $T_w^\star$ is the ratio of the actual wall temperature and the temperature of the molecules in the gas stream. The experimental data do show that the drag force depends linearly on $Kn^{-1}$ and that for the experimental hypersonic-flow condition $C_1/C_0$ is proportional to *S*. If we fit the experimental data for $T_W^\star = 0.70$, which is closest to the value $T_W^\star = 1$ considered in our paper, we find $C_1/C_0 = -0.634$ with S = 1.87 to be compared with $C_1/C_0 = -(0.62 \pm 0.03)\,S$ from Table 2. Although these two estimates do not correspond to the same wall temperatures, the similar magnitude is encouraging. For a more technical comparison one would need to evaluate the collision integrals for the experimental wall temperatures and with the hard-sphere molecular cross-section replaced with that for nitrogen molecules. This can be done by a straightforward generalization of the collision integrals presented in this paper.

Available theoretical estimates for the drag coefficient of a disc as a function of the inverse Knudsen number can be found in an article of Riganti [30]. Here we are specifically concerned with the initial dependence of the drag coefficient on $Kn^{-1}$. From Fig. 15 in [30] one sees that none of the available theoretical estimates reproduces the initial slope of the disc drag as a function of $Kn^{-1}$ as observed by Keel *et al.* [29], except for a model of Willis in which the molecular interaction is approximated by an empirical modified Krook model. However, even this model fails to account for the experimental data at a much lower wall temperature [29]. Hence, there is still a need for a more fundamental predictive theoretical approach as presented in the present paper.

## 6 Drag of a Sphere in Nearly Free Molecular Flow

6.1 Collision integrals for drag on a sphere

Next we consider the drag force on a sphere with a radius *R* small compared to the mean free path of the molecules in the gas. We can use the same dimensionless quantities that were introduced in Sect. 5.1 for evaluating the drag coefficient of a disc. The drag coefficient $C_D$ is again represented by (5.4) and (5.5), where *R* is now to be identified with the radius of the sphere. The drag coefficient $C_0$ in the free-molecular flow limit is now given by

$$C_0 = -\frac{2}{\pi S^2} \int d\mathbf{w}\, f^*(\mathbf{w};\mathbf{S}) \int_{\mathbf{w}\cdot\hat{\mathbf{R}}\leq 0} d\hat{\mathbf{R}}\, |\mathbf{w}\cdot\hat{\mathbf{n}}| \int d\mathbf{w}'\, \eta^*(\mathbf{w}')(\mathbf{w}'-\mathbf{w})\cdot\mathbf{S}. \qquad (6.1)$$

The contributions to the coefficient $C_1$ of the first inverse-Knudsen number correction

$$C_1 = C_{H1} + C_{H2} + 2(C_{R1} + C_{R2}) \qquad (6.2)$$

are again determined by the collision integrals (5.9) – (5.11), if $\mathbf{r}_1$ and $\hat{\mathbf{n}}_1$ are identified with the unit vector $\hat{\mathbf{R}}_1$ in the direction of $\mathbf{R}_1$ and $\mathbf{r}_2$ and $\hat{\mathbf{n}}_2$ with the unit vector $\hat{\mathbf{R}}_2$ in the direction of $\mathbf{R}_2$, with the location of $\mathbf{R}_1$ and $\mathbf{R}_2$ indicated in Figs. 3 and 5. The dimensionless velocities $\mathbf{w}_i, \mathbf{w}'_i, \mathbf{w}''_i, \mathbf{w}'''_i$ can be physically identified with the velocities $\mathbf{v}_i, \mathbf{v}'_i, \mathbf{v}''_i, \mathbf{v}'''_i$ in Figs. 3 and 5 with the radius of the sphere normalized to unity.

6.2 Drag coefficient of a sphere in the free-molecular-low limit

The expression (6.1) for $C_0$ can be readily evaluated analytically and one obtains

$$C_0(S) = \frac{(2S^2+1)}{\sqrt{\pi}S^3}e^{-S^2} + \frac{(4S^4+4S^2-1)}{2S^4}\text{erf}(S) + \frac{2\sqrt{\pi}}{3S}. \tag{6.3}$$

In the low- and high-speed limits the drag coefficient of a sphere reduces to

$$\lim_{S\to 0} C_0(S) = \frac{2\pi+16}{3\sqrt{\pi}S} \simeq \frac{4.191}{S}, \tag{6.4}$$

$$\lim_{S\to\infty} C_0(S) = 2. \tag{6.5}$$

The dependence of $C_0(S)$ of a sphere on the speed ratio $S$, shown in Fig. 9, is very similar to that of a disc, shown in Fig. 7.

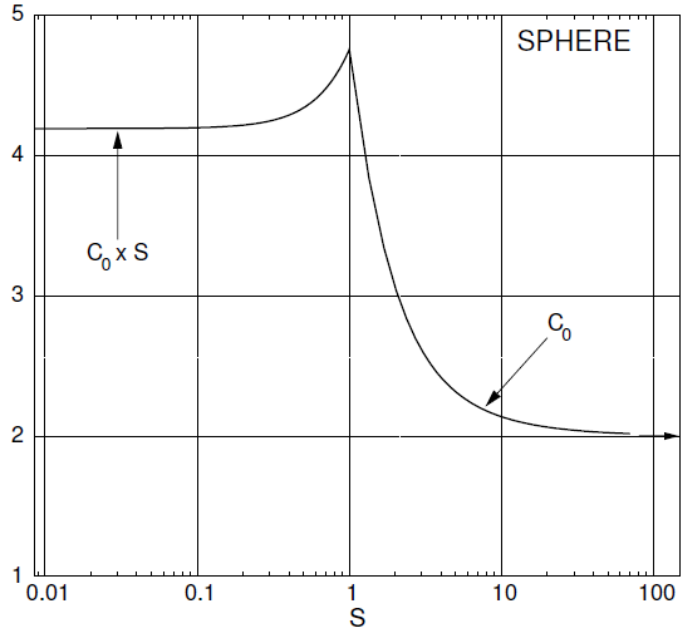

**Fig. 9** Drag coefficient of a sphere in free molecular flow as a function of the speed ratio $S$

6.3 Drag coefficient of a sphere in nearly free molecular flow

In the high-speed limit or beam limit, we can approximate the distribution functions $f^*(\mathbf{w}_i;S)$ by a delta function in accordance with (5.17). Again $\lim_{S\to\infty} C_{H2}(S)$ and $\lim_{S\to\infty} C_{R2}(S)$ do not contribute in this limit. As in the case of a disc, $\lim_{S\to\infty} C_{H1}(S)$ can again be evaluated analytically. However, to determine $\lim_{S\to\infty} C_{R1}(S)$ we can handle only some of the integrations in (5.11) analytically with the remainder to be estimated numerically. We then obtain

$$\lim_{S\to\infty} C_{H1}(S) = -\sqrt{\frac{\pi}{2}} S \approx -1.253 S \tag{6.6}$$

$$\lim_{S\to\infty} C_{R1}(S) = +(0.510 \pm 0.002) S, \tag{6.7}$$

so that

$$\lim_{S\to\infty} C_1(S) = \lim_{S\to\infty}\{C_{H1}(S) + 2 C_{R1}(S)\} = -(0.233 \pm 0.004) S. \tag{6.8}$$

In the low-speed limit we can again expand the Maxwell distributions in accordance with (5.21). Again after some simplifications, all collision integrals were evaluated numerically with the results presented in Table 3.

**Table 3** Coefficient $C_1$ of the first correction to the drag of a sphere in the low-speed limit

$$\begin{cases} \lim_{S\to 0} C_{H1}(S) = -(2.585 \pm 0.014) S^{-1} \\ \lim_{S\to 0} C_{H2}(S) = +(0.299 \pm 0.008) S^{-1} \\ \lim_{S\to 0} C_H(S) = \lim_{S\to 0}\{C_{H1}(S) + C_{H2}(S)\} = -(2.29 \pm 0.01) S^{-1} \end{cases}$$

$$\begin{cases} \lim_{S\to 0} C_{R1}(S) = +(0.59 \pm 0.03) S^{-1} \\ \lim_{S\to 0} C_{R2}(S) = -(0.059 \pm 0.007) S^{-1} \\ \lim_{S\to 0} C_R(S) = \lim_{S\to 0}\{C_{R1}(S) + C_{R2}(S)\} = +(0.53 \pm 0.03) S^{-1} \end{cases}$$

$$\lim_{S\to 0} C_1(S) = \lim_{S\to 0}\{C_H(S) + 2 C_R(S)\} = -(1.23 \pm 0.05) S^{-1}$$

To obtain the drag coefficient at arbitrary values of the speed ratio $S$, we need to retain the full expression (5.2) in the collision integrals (5.9) – (5.12). For various values of the speed ratio $S$, the collision integrals were again evaluated numerically. The results obtained by averaging over 200,000 random points are presented in Table 4. The dependence of $C_1/C_0$ on the speed ratio $S$ is shown in Fig. 10.

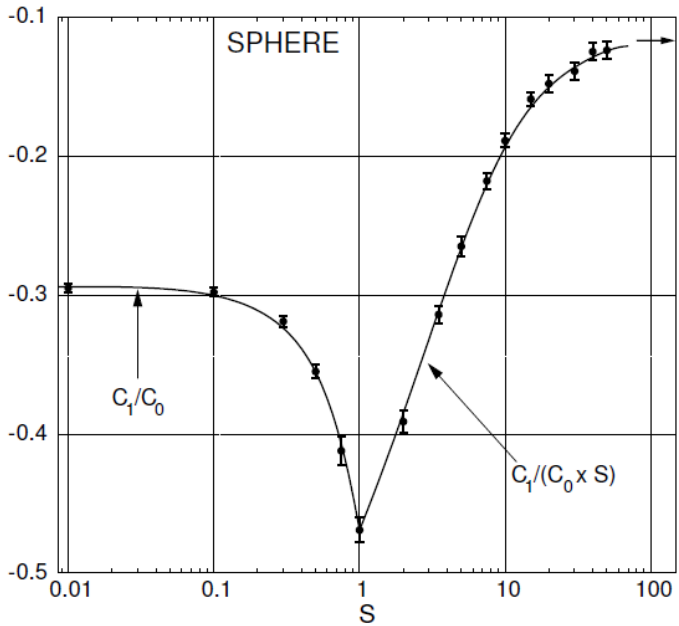

**Fig. 10** Coefficient $C_1/C_0$ of the first inverse-Knudsen number correction for a sphere as a function of the speed ratio $S$. Symbols with error bars represent actual numerical data; curves indicate a smooth interpolation of numerical data.

**Table 4** Drag coefficient of a sphere in the nearly free molecular flow regime as a function of the speed ratio $S$.

$$C_D = C_0\left[1+\frac{C_1}{C_0}Kn^{-1}\right] = C_0\left[1+\left(\frac{C_H+2C_R}{C_0}\right)Kn^{-1}\right]$$

| $S$ | $C_0$ | $C_H/C_0$ | $C_R/C_0$ | $C_1/C_0$ |
|---|---|---|---|---|
| 0    | $4.1906\ S^{-1}$ | $-0.544 \pm 0.009$ | $+0.124 \pm 0.001$ | $-0.297 \pm 0.003$ |
| 0.1  | $4.1967\ S^{-1}$ | $-0.549 \pm 0.001$ | $+0.125 \pm 0.001$ | $-0.298 \pm 0.003$ |
| 0.3  | $4.2445\ S^{-1}$ | $-0.586 \pm 0.001$ | $+0.134 \pm 0.002$ | $-0.319 \pm 0.004$ |
| 0.5  | $4.3385\ S^{-1}$ | $-0.656 \pm 0.002$ | $+0.151 \pm 0.003$ | $-0.355 \pm 0.005$ |
| 0.75 | $4.5164\ S^{-1}$ | $-0.780 \pm 0.005$ | $+0.184 \pm 0.005$ | $-0.412 \pm 0.010$ |
| 1.0  | $4.7538\ S^{-1}$ | $-0.914 \pm 0.007$ | $+0.222 \pm 0.003$ | $-0.469 \pm 0.009$ |
| 1.0  | 4.7538 | $-(0.914 \pm 0.007)\ S$ | $+(0.222 \pm 0.003)\ S$ | $-(0.469 \pm 0.009)S$ |
| 2.0  | 3.0596 | $-(0.815 \pm 0.007)\ S$ | $+(0.212 \pm 0.002)\ S$ | $-(0.391 \pm 0.008)S$ |
| 3.5  | 2.4975 | $-(0.751 \pm 0.005)\ S$ | $+(0.219 \pm 0.002)\ S$ | $-(0.314 \pm 0.006)S$ |
| 5.0  | 2.3155 | $-(0.721 \pm 0.006)\ S$ | $+(0.228 \pm 0.002)\ S$ | $-(0.265 \pm 0.007)S$ |
| 7.5  | 2.1929 | $-(0.690 \pm 0.004)\ S$ | $+(0.236 \pm 0.002)\ S$ | $-(0.218 \pm 0.006)S$ |
| 10   | 2.1381 | $-(0.668 \pm 0.003)\ S$ | $+(0.239 \pm 0.002)\ S$ | $-(0.189 \pm 0.005)S$ |
| 15   | 2.0877 | $-(0.652 \pm 0.003)\ S$ | $+(0.246 \pm 0.002)\ S$ | $-(0.159 \pm 0.005)S$ |
| 20   | 2.0641 | $-(0.647 \pm 0.004)\ S$ | $+(0.250 \pm 0.002)\ S$ | $-(0.148 \pm 0.006)S$ |
| 30   | 2.0416 | $-(0.643 \pm 0.004)\ S$ | $+(0.254 \pm 0.003)\ S$ | $-(0.139 \pm 0.006)S$ |
| 40   | 2.0308 | $-(0.635 \pm 0.003)\ S$ | $+(0.255 \pm 0.004)\ S$ | $-(0.125 \pm 0.006)S$ |
| 50   | 2.0244 | $-(0.633 \pm 0.004)\ S$ | $+(0.255 \pm 0.003)\ S$ | $-(0.124 \pm 0.006)S$ |
| $\infty$ | 2 | $-0.627\ S$ | $+(0.255 \pm 0.001)\ S$ | $-(0.117 \pm 0.002)S$ |

Note: $Kn^{-1} = \sqrt{2}\pi n\sigma^2 R$

6.4 Discussion

The drag exerted on a sphere in the nearly molecular-flow regime has been studied by a number of authors. Most of these studies are concerned either with low velocities $S \ll 1$ or large velocities $S \gg 1$. We shall discuss these two cases separately.

Our results are based on a Knudsen-number iterated solution of the Boltzmann equation for a gas of hard sphere molecules in the presence of an object. A study of the sphere drag, based on the Boltzmann equation for Maxwellian molecules, *i.e.*, for molecules that repel each other with forces proportional to the inverse fifth power of the intermolecular separation, has been reported by Liu *et al*. [31]. They expanded the collision integrals in terms of Hermite polynomials as functions of the molecular velocity; this procedure leads to an expansion of the drag force in which higher-order terms of the speed ratio $S$ are neglected. A comparison of our results with those of Liu *et al*. is presented in Table 5. For very small speed ratios $S$ the results are equal suggesting that the low-velocity drag is insensitive to the details of the intermolecular potential of the molecules in the gas. However, in contrast to the results of Liu *et al*., we predict that $C_1$ should increase with increasing $S$. This difference may be due to a failure of the expansion procedure of Liu *et al*. for larger values of $S$, the convergence of which has been questioned by Willis [32], or to the dependence of the bimolecular cross-section on the relative velocity of the colliding pair of molecules, characteristic for Maxwell molecules.

**Table 5** Comparison of our results for the sphere drag with those of Liu *et al*. [31]

| $S$ | Liu,Pang,Jew (Maxwellian Molecules) | This work (Hard Sphere Molecules) |
|---|---|---|
| 0 | $\frac{C_1}{C_0} = -0.298$ | $\frac{C_1}{C_0} = -0.297 \pm 0.003$ |
| 0.1 | $-0.300$ | $-0.298 \pm 0.003$ |
| 0.3 | $-0.302$ | $-0.319 \pm 0.004$ |
| 0.5 | $-0.308$ | $-0.355 \pm 0.005$ |
| 0.75 | $-0.310$ | $-0.412 \pm 0.010$ |
| 1.00 | $-0.296$ | $-0.469 \pm 0.009$ |

Rather than starting from the Boltzmann equation, investigators have tried to evaluate the drag force from the Bhatnagar, Gross, and Krook (BGK) model equation [6, 32]. Although judicious application of the BGK equation has yielded some encouraging results [33], its predictive power is limited [34]. From a Knudsen iteration procedure, based on the BGK equation for the drag coefficient of a sphere, Willis found $\lim_{S \to 0}(C_1/C_0) = -0.366$ [32] to be compared with our result $\lim_{S \to 0}(C_1/C_0) = -0.3$. This order of magnitude value seems to be in good agreement with results deduced from experiments of Millikan [35] for an oil drop in air. Differences are to be expected due to various collision cross-sections of the air molecules, and possible effects related to a proper description of air-oil drop interactions.

A survey of theoretical values reported for the correction to the drag of a sphere in nearly free molecular flow for large values of $S$ is presented in Table 6. Baker and Charwat [36] and Perepukhov [37] start from the same model considered by us, but then introduce a number of approximations. Hence, the extent to which their results differ from our value indicates the effect of their approximations. In particular the drastic approximations introduced by Baker and Charwat do not appear to be justified. The value obtained by Rose [38] is based on the BGK equation and the one obtained by Willis [39] on a modified BGK equation. As mentioned above, their validity can only be judged a posteriori from a comparison with the solution of the iterated Boltzmann equation.

**Table 6** Theoretical values reported for $\lim_{S \to \infty}(C_1/C_0)$ of a sphere

| | |
|---|---|
| $\lim_{S \to \infty}(C_1/C_0) = -0.24S,$ | Baker and Charwat [36] |
| $-0.143S$ | Perepukhov [37] |
| $-0.33S$ | Rose [38] |
| $-0.165S$ | Willis [39] |
| $-(0.117 \pm 0.002)S$ | This work |

The observation that the theoretical values of $C_1/C_0$ become proportional to the speed ratio $S$ for large velocities deserves some further comments. The expansion (1.3) is only valid for $Kn^{-1} \ll 1$. However, the parameter $Kn^{-1} = \sqrt{2}\pi n\sigma^2 R$ in the expansion for the drag coefficient represents the inverse Knudsen number in the absence of the object. In order for the expansion to be valid we must require that the local inverse Knudsen number near the object is substantially smaller than unity. When we denote the local Knudsen number by $Kn_{eff}$, then at small velocities $Kn_{eff}^{-1} \simeq Kn^{-1}$. However, at large velocities $Kn_{eff}^{-1} \simeq SKn^{-1} = S\sqrt{2}\pi n\sigma^2 R$, since at large velocities the mean free path of the molecules emitted from the surface becomes inversely proportional to the speed ratio $S$. That is, if we rewrite expansion (5.16) in terms of the local Knudsen number

$$C_D = C_0 \left[ 1 + \frac{C_{1,eff}}{C_0} Kn_{eff}^{-1} \right], \tag{6.9}$$

subject to the condition $Kn_{eff}^{-1} \ll 1$, then $\lim_{S \to \infty}(C_{1,eff}/C_0) = -0.117$ becomes independent of the speed ratio $S$. Thus the apparent increase of $C_1/C_0$ at large velocities $S$ is a consequence of the fact that the local inverse Knudsen number itself increases with $S$. The condition $Kn_{eff}^{-1} \ll 1$ implies that for large values of the speed ratio $Kn^{-1} \ll S^{-1}$, so that the range of validity of the expansion decreases with increasing value of $S$. Of course, this limitation applies to all theoretical results obtained by a Knudsen-number-iteration procedure.

## 7 Discussion

In this paper we have derived an expression for the drag coefficient in nearly free molecular flow of the form

$$C_D = C_0 + C_1 Kn^{-1}. \tag{7.1}$$

The coefficient $C_0$ represents the drag coefficient in the limit of free molecular flow. We have shown that the coefficient $C_1$ of the first inverse Knudsen number correction is determined by a set of well-defined collision integrals associated with dynamical events involving two gas molecules and the object. These collision integrals can be formulated for objects of any shape. To demonstrate the feasibility of the theoretical procedure for applications, we have evaluated these collision integrals for the coefficient $C_1$ of a disc and a sphere as a function of the speed ratio $S$ assuming that the gas molecules are re-emitted diffusively by the object. For convenience, we took the temperature of the object to be equal to the temperature of the gas stream and assumed that the molecules of the gas could be treated as hard spheres. These approximations are not essential. Other temperatures of the object and molecules with more complicated interaction potentials can be handled by inserting the appropriate binary collision operators in the expressions derived for the collision integrals. The theoretical results presented in this paper may be used as a check of the quality of approximate theories for the first inverse-Knudsen number correction to the drag coefficients of objects in nearly free molecular flow.

**Acknowledgments** We acknowledge a fruitful collaboration with W.A. Kuperman in an early stage of this research.

**References**

1. S.A. Schaaf, in S.S. Flügge, (Ed.) Handbuch der Physik, VIII, part 2, Springer, Berlin, 1963, 591-624.
2. E.P. Munz, Ann. Rev. Fluid Mech. 21 (1989) 387-417.
3. Y. Sone, Molecular Gas Dynamics: Theory, Techniques, and Applications, Birkhäuser, Boston, 2007.
4. F.S. Sherman, Ann. Rev. Fluid Mech. 1 (1969) 317-340.
5. M.N. Kagan, Rarefied Gas Dynamics, Plenum Press, New York, 1969.
6. C. Cercignani, Theory and Application of the Boltzmann Equation, Elsevier, New York, 1987.
7. C. Shen, Rarefied Gas Dynamics: Fundamentals, Simulations, and Micro Flows, Springer, Berlin, 2005.
8. H. van Beijeren, J.R. Dorfman, J. Stat. Phys. 23 (1980) 335-402.
9. H. van Beijeren, J.R. Dorfman, J. Stat. Phys. 23 (1980) 443-461.
10. C.B. Henderson, AIAA J. 14 (1976) 707-708.

11. J.R. Dorfman, W.A Kuperman, J.V. Sengers, C.F. McClure, Phys. Fluids 16 (1973) 2347-2349.
12. J.R. Dorfman, H. van Beijeren, C.F.McClure, Arch. Mech. Stos. 28 (1976) 333-352.
13. J.R. Dorfman, J.V. Sengers, C.F. McClure, Physica A 134 (1986) 283-322.
14. J.R. Dorfman, T.R. Kirkpatrick, J.V. Sengers, J.V, Ann. Rev. Phys. Chem. 45 (1994) 213-239.
15. M. Lunc, J. Lubonski, Arch. Mech. Stos. 8 (1956) 597-616.
16. W.A. Kuperman, Aerodynamic Forces on Objects in the Nearly Free Molecular Flow Regime, Ph.D. Thesis, Institute for Molecular Physics, University of Maryland, College Park, MD, 1972.
17. Y. Pomeau, Phys. Fluids 18 (1975) 277-288.
18. J.V. Sengers, Phys. Fluids 9 (1966) 1685-1696.
19. Y. Kan, J.R. Dorfman, Phys. Rev. A 16 (1977) 2447-2469.
20. T. Munakata, Progr. Theor. Phys. 58 (1977) 59-76.
21. J.V. Sengers, M.H. Ernst, D.T. Gillespie, J. Chem. Phys. 56 (1972) 5583-5601.
22. J.V. Sengers, D.T. Gillespie, J.J. Perez-Esandi, Physica A 96 (1978) 365-409.
23. M.H. Ernst, J.R. Dorfman, W.R. Hoegy, J.M.J. van Leeuwen, Physica 45 (1969) 127-146.
24. W.R. Hoegy, J.V. Sengers, Phys. Rev. A 2 (1970) 2461-2471.
25. J.R. Dorfman, M.H. Ernst, J. Stat. Phys. 57 (1989) 581-593.
26. Y.-Y. Lin Wang, J.V. Sengers, Drag Force on Objects in the Nearly Free Molecular Flow Regime as a Function of Speed Ratio, Technical Report AEDC-TR-74-79, Arnold Engineering Development Center, Arnold Air Force Station, TN, 1974, 100pp.
27. H. Ashley, J. Aeronaut. Sci. 16 (1949) 95-104.
28. D.R. Willis, Bull. Am. Phys. Soc. 16 (1971) 1333.
29. A.G. Keel, T.E. Chamberlain, D.R. Willis, in: K. Karamcheti, (Ed.) Rarefied Gas Dynamics, Academic Press, New York, 1974, 327-333.
30. R. Riganti, Meccanica 9 (1974) 80-87.
31. V.C. Liu, S.C. Pang, H. Jew, Phys. Fluids 8 (1965) 788-796.
32. D.R. Willis, Phys. Fluids 9 (1966) 2522-2524.
33. F.S. Sherman, D.R. Willis, G.J. Maslach, in H. Görtler, (Ed.), Applied Mechanics, Springer, Berlin, 1966, 871-877.
34. D.R. Willis, in I.L. Trilling, H.Y. Wachman, (Eds.), Rarefied Gas Dynamics, Advances in Applied Mechanics, Suppl. 5, Vol. 1, Academic Press, New York, 1969, 61-70.
35. R.A. Millikan, Phys. Rev. 22 (1923) 1-23.
36. R.M.L. Baker, A.F. Charwat, Phys. Fluids 1 (1958) 73-81.
37. V.A. Perepukhov, USSR Computational Mathematics and Mathematical Physics 7 (1967) 276-288.
38. M.H. Rose, Phys. Fluids 7 (1964) 1262-1269.
39. G.J. Maslach, D.R. Willis, S. Tang, D. Ko, in J.H. de Leeuw (Ed), Rarefied Gas Dynamics, Advances in Applied Mechanics, Suppl. 3, Vol. I, Academic Press, New York, 1965, 433-44